\documentclass[twocolumn,twocolappendix]{aastex631}
\usepackage{float}
\usepackage{xspace}
\usepackage{amsmath}
\usepackage{hyperref}    
\usepackage{xcolor}

\newcommand{\OIII}{[\ion{O}{3}]\xspace}

\received{}
\revised{}
\accepted{}
\submitjournal{ApJ}

\shorttitle{PNLF Bright-End Cutoff}
\shortauthors{Jacoby \& Ciardullo}

\graphicspath{{./}{figures/}}

\begin{document}

\title{The Critical Role of Dust On The \OIII Planetary Nebula Luminosity Function's Bright-End Cutoff}

\correspondingauthor{George H. Jacoby}
\email{george.jacoby@noirlab.edu}

\author[0000-0001-7970-0277]{George H. Jacoby}   
\affiliation{NSF's NOIRLab, 950 N. Cherry Ave., Tucson, AZ 85719, USA}

\author[0000-0002-1328-0211]{Robin Ciardullo}   
\affiliation{Department of Astronomy \& Astrophysics, The Pennsylvania State University, University Park, PA 16802, USA}
\affiliation{Institute for Gravitation and the Cosmos, The Pennsylvania State University, University Park, PA 16802}

\begin{abstract}

We examine the relationship between circumnebular extinction and core mass for sets of \OIII-bright PNe in the Large Magellanic Cloud and M31.  We confirm that for PNe within one magnitude of the Planetary Nebula Luminosity Function's (PNLF's) bright-end cutoff magnitude ($M^*$), higher core-mass PNe are disproportionally affected by greater circumnebular extinction.  We show that this result can explain why the PNLF cutoff is so insensitive to population age. In younger populations, the higher-mass, higher-luminosity cores experience greater circumnebular extinction from the dust created by their AGB progenitors compared to the lower-mass cores.  We further show that when our core-mass-nebular extinction law is combined with post-AGB stellar evolutionary models, the result is a large range of population ages where the brightest PNe all have nearly identical \OIII luminosities.  Finally, we note that while there is some uncertainty about whether the oldest stellar populations can produce planetary nebulae (PNe) as bright as $M^*$, this issue is resolved if the initial-final mass relation (IFMR) for the lowest-mass stars results in slightly more massive cores, as observed in some clusters. Alternatively, introducing a small amount of intrinsic scatter ($0.022 M_\odot$) into the IFMR also addresses this uncertainty.

\end{abstract}

\section{Introduction} 
\label{sec:intro}
Since the 1990s, the bright-end cutoff of the \OIII $\lambda 5007$ Planetary Nebula Luminosity Function (PNLF) has been used as a quality extragalactic distance indicator for large galaxies of all Hubble types.  Tests within galaxies, galaxy groups, and galaxy clusters have shown that the absolute luminosity of the PNLF cutoff is very nearly constant across a wide range of stellar populations \citep[e.g.,][]{Jacoby1997, Ciardullo+02}, with the brightest PNe emitting $\sim 620 L_{\odot}$ in the \OIII $\lambda 5007$ emission line.  Moreover, as long as the oxygen abundance of a system is greater than $12 + \log \textrm{O/H} \sim 8.3$, this luminosity, which corresponds to an absolute $\lambda 5007$ magnitude of $M^* \approx -4.5$ \citep{Jacoby1989}, has little if any, dependence on population age or metallicity \citep[e.g.,][]{Ciardullo+02, Ciardullo2022}.  But why should the PNLF have a sharp bright-end cutoff, and what causes the cutoff's absolute magnitude to have so little dependence on stellar population?

The rationale for the existence of a bright-end cutoff was first discussed by \citet{Jacoby1989}, who argued that a narrow distribution of central star masses, combined with the rapid evolution of more massive cores, would moderate \OIII luminosities.  But the validity of this argument is limited: while it is true that the mass distribution of Milky Way central stars is sharply peaked \citep[e.g.,][]{Tremblay+16, Torres+21}, the location of that peak should be sensitive to population age, in accordance with the initial-final mass relation \citep[IFMR; e.g.,][]{Cummings+18, ElBadry+18}.  This fact was vividly demonstrated by \citet{Marigo+04}, whose models showed that the absolute luminosity for the most luminous PNe of a $\sim 1$~Gyr stellar population should be $\sim 3$~mag brighter than that for similar PNe in a $\sim 10$~Gyr system.  Furthermore, these same calculations showed that the PN luminosities predicted for old stellar populations were far fainter than those of the PNe observed in the elliptical galaxies of the Leo~I, Virgo, and Fornax clusters.

To help explain the presence of $M_{5007} \sim -4.5$ PNe in old populations, \citet{Ciardullo+05} proposed a model where the products of binary star mergers formed from conservative mass transfer are responsible for populating the bright end of the PNLF\null. While such a scenario did indeed increase the core masses and luminosities of the brightest PNe in a galaxy, it did not address the more significant issue of the uniform shape of the PNLF's cutoff or the apparent hard upper limit on the amount of \OIII luminosity that can be produced by a PN.

What limits the \OIII $\lambda 5007$ luminosities of PNe to $\sim 620 L_{\odot}$?   One possibility that has not been studied in detail is the attenuation caused by the dust that was created by the PN's progenitor star when it was on the asymptotic giant branch (AGB)\null.   For all but the lowest-mass objects, this dust should still be in the vicinity of the post-AGB star when the core becomes hot enough to ionize hydrogen and oxygen. In principle, this self-extinction can greatly affect the observed luminosity of an \OIII-bright planetary. Moreover, since the envelope mass ejected by an AGB star \citep[e.g.,][]{Cummings+18} along with the mass of dust created \citep{Ventura+14}, and the rate at which a post-AGB star evolves across the HR diagram are both strongly dependent on stellar mass \citep[e.g.,][]{Blocker+95, Vassiliadis+94, Miller-Bertolami2016}, it is easy to imagine models in which the amount of self-extinction affecting a PN increases rapidly with the mass and luminosity of its central star.

We know that the extinction surrounding bright extragalactic PNe is not insignificant.  For example, \citet{Herrmann+09a} noted that in M94, M33, and the LMC, the PNe within $\sim 2$~mag of the PNLF cutoff have, on average, $A_V \sim 0.6$ more extinction than their immediate surroundings.  Spectrophotometric analyses of PNe in M31 have produced a similar result \citep[e.g.,][]{Jacoby+99, Kwitter+12, Fang+18, Ueta+22}.  In fact, by measuring the circumnebular extinction of 23 PNe projected within the bulge of M31 (all within the top $\sim 1$~mag of the PNLF) and comparing these data to similar archival measurements of PNe in the LMC \citep{Reid+10}, NGC\,4697 \citep{Mendez+05}, and NGC\,5128 \citep{Walsh+12}, \citet{Davis+18} concluded that the intrinsic (dereddened) shape of the PNLF cutoff is likely quite different (and much brighter) than that of the observed luminosity function.  These data show that more than 75\% of \OIII bright PNe have $A_{5007} > 0.5$~mag of circumnebular extinction ($c_{{\rm H}\beta} > 0.23$), and $\sim 25\%$ have $A_{5007} > 1.0$~mag ($c_{{\rm H}\beta} > 0.46$).

The idea that circumnebular dust is the mechanism responsible for truncating the bright end of the PNLF was first suggested by \citet{Ciardullo+99}.  These authors noted that in the LMC, there appeared to be a correlation between an object's nebular reddening and its core mass and, therefore, the mass and age of its progenitor star. The effects of dust extinction on the PNLF were further demonstrated in Figure~2 of \citet{Ciardullo2012}, and a simplified prescription was incorporated into the models of \citet{Gesicki+18} and \citet{Yao+23}. 

In this paper, we update the efforts of \citet{Ciardullo+99} within the framework of the newer \citet{Miller-Bertolami2016} stellar evolution models and adopt the extinction versus core mass relations observed for \OIII-bright PNe in the LMC and M31. By doing so, we demonstrate the viability of a straightforward ``proof-of-concept'' model, in which circumnebular dust truncates the bright end of the \OIII PNLF to the same absolute magnitude over a wide range of stellar masses.  

Section~\S\ref{sec:data} summarizes the sources for the data used in this study. In \S\ref{sec:approach}, we describe the procedures used for simulating the maximum observed luminosity for PNe across a wide range of population ages. {  In \S\ref{sec:extinction-coremass}, we explore the details of the correlation between core masses and extinction and show that PNe with the highest-mass central stars generally do not populate the top magnitude of the PNLF\null.} In \S\ref{sec:results}, we illustrate the uniformity of $M_{5007}$ maximum magnitudes as a function of progenitor mass for each of two sets of data:  one based on the analyses of LMC PNe and the other derived from M31 PN observations. In \S\ref{sec:discussion} we describe the successes and limitations of the proof-of-concept model, and in \S\ref{sec:conclusions}, we review the implications of the results.

\section{The Data}
\label{sec:data}

To analyze the effect of dust on the bright end of the PNLF, we need samples of PNe with reliable absolute \OIII magnitudes, known nebular extinctions, and well-determined bolometric luminosities for their central stars. While the \textit{Gaia} satellite is revolutionizing our knowledge of PN distances within the Milky Way \citep[e.g.,][]{Gonzalez-Santamaria+21, Bucciarelli+23}, very few well-measured Galactic PNe have \OIII absolute magnitudes that place them in the top $\sim 1$~mag of the PNLF \citep{Chornay+23}, and fewer yet have robust measures of their central stars' positions on the HR diagram or their amount of circumnebular reddening.  Hence, for this study, we use the PNe of nearby galaxies, where the distances are well-known, and the effects of circumnebular dust can be disentangled from the foreground reddening of the Milky Way.

{ 
\subsection{The LMC and M31 Samples}
}

Our study uses subsets of Large Magellanic Cloud and M31 PNe, whose nebular reddening and central star properties (bolometric luminosity and effective temperature) are relatively well-determined via photoionization models.  {  To ensure self-consistency, we chose for analysis the largest and most uniform PN samples available. Although measurements exist from several smaller investigations, differences in survey depth, resolution, wavelength coverage, and reduction/analysis techniques would likely introduce unquantifiable systematics into our results.  The samples described below therefore represent our attempt to achieve maximum uniformity over a sample sized large enough to draw robust conclusions.}

Our LMC data come solely from the PN analyses of \citet{Dopita+91a, Dopita+91b} and \citet{Dopita+97}, with the former two references obtaining their spectrophotometry (and nebular extinctions) from \citet{Meatheringham+91a, Meatheringham+91b}.  Since we are only concerned with the PNe that define the bright-end cutoff of the PNLF, we use the PNe's observed \OIII $\lambda 5007$ magnitudes as measured by \citet{Jacoby+90b} to restrict our analysis to objects within one magnitude of $M^*$.  Fourteen PNe satisfy these criteria; these are listed in Table~\ref{tab:pnlist}.

To explore a different PN population, we use similar measurements of M31 PNe from the spectrophotometric analyses of \citet{Kwitter+12}, \citet{Balick+13}, \citet{Fang+18}, and \citet{Ueta+22}.  (The latter is a reanalysis of data originally presented in \citealp{Galera+22}.) These works, which are all based on 8-m class telescope spectrophotometry, provide extinctions and central star properties for 28 PNe with \OIII luminosities within one magnitude of $M^*$.  (Three additional \OIII-bright bulge PNe are available from the 4-m observations of \citet{Jacoby+99}; however, since the errors on their extinction measurements are 3 to 5 times larger than those from the other studies, these objects have been excluded from the analysis.)   Additionally, four of the \citet{Kwitter+12} PNe have also been observed and analyzed independently by \citet{Galera+22} and \citet{Ueta+22}.  These repeat measurements provide some information about the uncertainties associated with the determinations of PN core mass and nebular extinction. The M31 PNe used in our analysis are also given in Table~\ref{tab:pnlist}.

We note that the sets of PNe being analyzed do not constitute statistically complete samples.  However,  they are representative of the bright PNe present in both galaxies.  In the LMC, the dataset contains 14 of the brightest 23 PNe in the galaxy, and an \citet{AD-Test} test confirms that the magnitude distribution of the objects in our sample is consistent with that of the full population of PNe within one magnitude of $M^*$. Similar agreement is found between the M31 PNe being analyzed and the full sample of $M_{5007} < M^* + 1$ PNe cataloged by \citet{Merrett+06}.  Thus, the results found in our study should apply to the overall population of [\ion{O}{3}]-bright PNe in the galaxies.

\begin{deluxetable*}{lccccDccl}
\tabletypesize{\footnotesize} 
\tablecaption{List of Analyzed PNe
\label{tab:pnlist} } 
\tablehead{
 &      &\multicolumn{2}{c}{Central Star}  &\colhead{Core Mass} &\multicolumn{3}{c}{Nebular Extinction} \\ [-6pt]
\colhead{ID\tablenotemark{\scriptsize a}} &\colhead{$M_{5007}$\tablenotemark{\scriptsize b}}   &\colhead{Log $L/L_{\odot}$} &\colhead{Log $T_{\rm eff}$} &\colhead{($M/M_{\odot})$\tablenotemark{\scriptsize c}} &\multicolumn2c{$c_{{\rm H}\beta}$\tablenotemark{\scriptsize d}} &\colhead{$\sigma(c_{{\rm H}\beta})$\tablenotemark{\scriptsize e}} &\colhead{12 + $\log$ O/H} &\colhead{Source} }
\decimals
\startdata
\multicolumn{8}{l}{\textit{LMC Objects}} \\
SMP 1   & $-3.67$  & 3.845  & 4.819  & 0.582  & 0.057    &\dots   &8.301  & \citet{Dopita+91a} \\
SMP 6   & $-3.59$  & 3.916  & 5.176  & 0.630  & 0.313    &\dots   &8.342  & \citet{Dopita+91a} \\
SMP 15  & $-3.76$  & 3.832  & 5.130  & 0.597  & 0.391    &\dots   &8.301  & \citet{Dopita+91a} \\
SMP 38  & $-3.81$  & 3.526  & 4.978  & 0.548  & 0.091    &\dots   &8.301  & \citet{Dopita+91b} \\
SMP 47  & $-3.86$  & 3.677  & 5.149  & 0.583  & $-$0.048 &\dots   &8.360  & \citet{Dopita+97} \\
SMP 61  & $-3.70$  & 3.73   & 4.841  & 0.567  & 0.091    &\dots   &8.322  & \citet{Dopita+97} \\
SMP 62  & $-4.39$  & 3.73   & 4.653  & 0.563  & 0.091    &\dots   &8.230  & \citet{Dopita+91b} \\
SMP 63  & $-3.91$  & 3.619  & 4.914  & 0.553  & 0.113    &\dots   &8.279  & \citet{Dopita+97} \\
SMP 73  & $-4.14$  & 3.852  & 5.130  & 0.602  & 0.268    &\dots   &8.415  & \citet{Dopita+91b} \\
SMP 74  & $-3.61$  & 3.850  & 5.068  & 0.594  & 0.435    &\dots   &8.041  & \citet{Dopita+91b} \\
SMP 78  & $-4.04$  & 3.781  & 5.114  & 0.585  & 0.257    &\dots   &8.398  & \citet{Dopita+91b} \\
SMP 89  & $-3.89$  & 3.932  & 4.996  & 0.621  & 0.557    &\dots   &8.380  & \citet{Dopita+97} \\
SMP 92  & $-4.00$  & 3.960  & 5.152  & 0.644  & 0.357    &\dots   &8.431  & \citet{Dopita+91b} \\
SMP 99  & $-4.10$  & 3.756  & 5.093  & 0.582  & 0.235    &\dots   &8.415  & \citet{Dopita+91b} \\
\multicolumn{8}{l}{\textit{M31 Objects}} \\
M  50   & $-$4.33  & 3.88   & 5.08   & 0.606  &  0.192  & 0.02   & 8.59   & \citet{Ueta+22} \\
M  71   & $-$3.83  & 3.58   & 5.05   & 0.562  &  0.162  & \dots  & 8.584  & \citet{Kwitter+12} \\
M 174   & $-$3.77  & 3.70   & 4.85   & 0.565  & -0.028  & 0.030  & 8.593  & \citet{Balick+13} \\
M 337   & $-$4.08  & 3.71   & 5.04   & 0.577  &  0.222  & \dots  & 8.698  & \citet{Kwitter+12} \\
M 553   & $-$4.12  & 3.60   & 5.02   & 0.561  &  0.052  & \dots  & 8.643  & \citet{Kwitter+12} \\
M 746   & $-$4.21  & 3.56   & 5.08   & 0.564  &  0.022  & \dots  & 8.704  & \citet{Kwitter+12} \\
M 795   & $-$4.10  & 3.70   & 5.03   & 0.576  &  0.222  & \dots  & 8.834  & \citet{Kwitter+12} \\
M 1074  & $-$3.78  & 3.68   & 4.99   & 0.570  &  0.082  & \dots  & 8.365  & \citet{Kwitter+12} \\
M 1074  & $-$3.78  & 3.76   & 5.00   & 0.578  &  0.122  & 0.02   & 8.47   & \citet{Ueta+22} \\
M 1583  & $-$3.95  & 3.65   & 5.15   & 0.578  &  0.232  & \dots  & 8.595  & \citet{Kwitter+12} \\
M 1596  & $-$3.95  & 3.46   & 5.06   & 0.551  &  0.022  & \dots  & 8.770  & \citet{Kwitter+12} \\
M 1596  & $-$3.95  & 3.75   & 5.11   & 0.579  &  0.112  & 0.02   & 8.63   & \citet{Ueta+22} \\
M 1675  & $-$3.99  & 4.05   & 5.05   & 0.679  &  0.452  & 0.05   & 8.65   & \citet{Ueta+22} \\
M 1687  & $-$4.50  & 3.87   & 5.01   & 0.596  &  0.152  & 0.02   & 8.49   & \citet{Ueta+22} \\
M 1985  & $-$3.91  & 3.84   & 5.05   & 0.591  &  0.322  & \dots  & 8.403  & \citet{Kwitter+12} \\
M 2068  & $-$4.47  & 4.04   & 5.01   & 0.672  &  0.282  & 0.03   & 8.60   & \citet{Ueta+22} \\
M 2240  & $-$3.96  & 3.68   & 5.08   & 0.577  &  0.222  & \dots  & 8.647  & \citet{Kwitter+12} \\
M 2432  & $-$3.97  & 3.665  & 5.137  & 0.578  &  0.157  & 0.040  & 8.573  & \citet{Fang+18} \\
M 2449  & $-$3.78  & 3.501  & 5.073  & 0.556  &  0.089  & 0.045  & 8.658  & \citet{Fang+18} \\
M 2471  & $-$3.97  & 3.35   & 5.14   & 0.561  &  0.012  & \dots  & 8.654  & \citet{Kwitter+12} \\
M 2471  & $-$3.97  & 3.82   & 5.10   & 0.591  &  0.252  & 0.04   & 8.53   & \citet{Ueta+22} \\
M 2472  & $-$3.97  & 3.58   & 5.00   & 0.556  &  0.062  & \dots  & 8.537  & \citet{Kwitter+12} \\
M 2496  & $-$4.24  & 4.00   & 4.86   & 0.646  &  0.052  & 0.031  & 8.425  & \citet{Balick+13} \\
M 2512  & $-$3.56\tablenotemark{\scriptsize f}  & 3.380  & 5.067  & 0.546  &  0.134  & 0.036  & 8.821  & \citet{Fang+18} \\
M 2538  & $-$4.41  & 4.03   & 5.07   & 0.670  &  0.262  & 0.10   & 8.38   & \citet{Ueta+22} \\
M 2624  & $-$3.94  & 3.64   & 5.26   & 0.597  &  0.122  & \dots  & 8.875  & \citet{Kwitter+12} \\
M 2690  & $-$3.89  & 3.65   & 5.01   & 0.567  &  0.072  & \dots  & 8.583  & \citet{Kwitter+12} \\
M 2694  & $-$4.06  & 3.85   & 5.03   & 0.592  &  0.272  & \dots  & 8.464  & \citet{Kwitter+12} \\
M 2860  & $-$3.82  & 3.47   & 5.06   & 0.552  &  0.062  & \dots  & 8.723  & \citet{Kwitter+12} \\
M 2860  & $-$3.82  & 3.97   & 5.03   & 0.640  &  0.452  & 0.08   & 8.65   & \citet{Ueta+22} \\
M 2895  & $-$3.88  & 3.331  & 5.240  & 0.578  & $-$0.001  & 0.042  & 8.823  & \citet{Fang+18} \\
Y 17    & $-$3.92  & 3.687  & 4.980  & 0.570  &  0.155  & 0.037  & 8.462  & \citet{Fang+18} \\
\enddata
\tablenotetext{a}{SMP: \citet{SMP78};  M: \citet{Merrett+06}; Y: \citet{Yuan+10} }
\tablenotetext{b}{Assuming $(m-M)_{5007} = 18.80$ for the LMC and 24.66 for M31}
\tablenotetext{c}{Found by interpolating $\log L$, $\log T_{\rm eff}$ in the \citet{Miller-Bertolami2016} grid of post-AGB evolutionary tracks}
\tablenotetext{d}{Assuming foreground $E(B-V) = 0.080$ for the LMC and $E(B-V) = 0.062$ for M31 }
\tablenotetext{e}{Quoted by \citet{Ueta+22}, otherwise derived from error propagation of Balmer-line strengths.}
{ \tablenotetext{f}{Magnitude in \citet{Fang+18} was based on the value in \citet{Yuan+10}}}
\end{deluxetable*}

{  
\subsection{Removal of Foreground Extinction
\label{subsec:extinction}}

The $c_{{\rm H}\beta}$ values measured for both the LMC and M31 PNe reflect the total amount of extinction along the line-of-sight.  This extinction comes from three components:  foreground dust in the Milky Way, foreground dust within the program galaxies themselves, and circumnebular dust ejected by the PN's progenitor.   

To remove the first component, we used the extinction maps of \citet{Schlegel+98} to derive $E(B-V) = 0.080$ for the LMC and $E(B-V) = 0.062$ for M31.  These reddenings are supported by the observed $c_{{\rm H}\beta}$ values of the galaxies' PNe; when this component is removed, the floor of the two reddening distributions is close to zero.

Quantifying the amount of extinction caused by foreground dust within the target galaxies is more problematic.  Yet it must be carefully considered:  if this internal component is larger for PNe from high-mass progenitors, it could introduce a systematic error into our measurement of the extinction versus core-mass relation.  To examine this possibility, we used the 3-D distribution of dust within the Milky Way, as measured by \citet{Guo+21}.  This model describes the vertical structure of dust in the solar neighborhood, i.e., at galactocentric radii between $\sim 6$ and $\sim 10$~kpc.  If we assume that the dust distribution in the target galaxies is similar, then the model predicts that the maximum amount of foreground extinction a PNe could experience is $A_{5007} \sim 0.15 \sec i$, where $i$ is the inclination of the galaxy.

Next, we assume that the vertical distributions of our program PNe are related to the ages of their progenitor stars. \citet{Mazzi+24} have used Gaia data to quantify the age-scale height relationship for Milky Way objects in the solar cylinder, and we adopt this same relation for our target galaxies.  

Finally, we apply these assumptions to the target galaxies and estimate the amount of foreground dust that might attenuate PNe of a given age (or, equivalently, progenitor mass).   This allows us to address the question of whether the foreground reddening of PNe observed in the top 1~mag of the PNLF can introduce a systematic term into our analysis of circumnebular extinction. 

The results of our model show that the effect of galactic internal extinction can vary greatly from object to object, with a distribution that depends strongly on the age of the target population.  For example, if a PN sub-population has a scale-height similar to that of the extinction, the distribution of their $A_{5007}$ values is roughly Gaussian in nature, with most objects having $A_{5007}$ values near the distribution's mean.  In contrast, if the scale height is much larger than the dust, the distribution of $A_{5007}$ values is almost flat and observations will record many more high- and low-extinction objects.  

Yet, almost paradoxically, this scale-height dependence has almost no effect on the mean properties of the observed sample: unless a galaxy is seen edge-on ($i > 80^\circ$), most of the PNe with \OIII absolute magnitudes in the top 1~mag of the PNLF will still be in the top 1~mag even after being reddened by foreground dust.  Moreover, because of the symmetry of the problem, the mean extinction affecting PNe from high-mass progenitors is not significantly different from that of PNe evolved from $\sim 1 M_{\odot}$ stars.  

Obviously, this analysis is quite simplified, as it assumes the vertical distributions of stars and dust in our program galaxies are the same as those measured for the solar neighborhood.  This will not always be true:  for instance, the projected dust distribution in M31's disk is quite complex, and the galaxy is known to have a ring of enhanced extinction ($A_V \sim 3$~mag) $\sim 9$~kpc from its center \citep{Dalcanton+15, Blana+18}. However, all but four of the M31 PNe listed in Table~\ref{tab:pnlist} are well beyond this feature (the median projected galactocentric radius of the sample is over 17~kpc);  at such large distances, M31's observed extinction is relatively low.  Additionally, \citet{Bhattacharya+19} have shown that the observed Balmer decrements of 413 of M31's PNe do not correlate with the position of the aforementioned dust ring or any other feature in the galaxy's line-of-sight extinction map.  Again, this supports the argument that internal extinction will not bias our results.

Our analysis implies that as long as the total extinction in a sight line through a galaxy at the position of its PNe does not become greater than $A_{5007} \sim 1$~mag, the effect of foreground dust should be primarily random in nature; systematic shifts with stellar population should be below the level at which a bias is introduced into our analysis.  With an inclination of $i \sim 35^\circ$, the LMC's expected internal extinction of $A_{5007} \lesssim 0.2$~mag falls well below this threshold ($c_{{\rm H}\beta} \sim 0.08$).  Due to its more extreme inclination ($i \sim 77^\circ$), the effect of internal extinction on M31's PNe is more serious, but even then,  we expect $A_{5007} \lesssim 0.7$~mag ($c_{{\rm H}\beta} \lesssim 0.28$) for our outer disk sample of objects.

Since the floor in the distribution of PN extinctions appears close to that found from Milky Way extinction alone, we do not correct our measurements for unrelated foreground extinction within the program galaxies.   Instead, we simply treat the component as a source of random scatter in the data.  
}

\section{The Effects of Extinction on the PNLF}
\label{sec:approach}

Our analysis of the effect of circumnebular dust on the PNLF does not attempt to model the creation and evolution of PNe, which is a time-dependent problem with many variables. Instead, we simply derive an empirical relationship between extinction and central star mass and use that to determine the effect that dust may have on a population's brightest planetaries.  The advantages of this approach are that our basic assumptions have no large uncertainties, and our analysis is tied directly to observational data.  The drawback is that the time dependencies associated with the evolution of the central star, its nebula, and the formation, destruction, and distribution of dust are ignored. However, because we are not trying to simulate the shape of the full PNLF but only quantify the location of the function's cutoff (i.e., the maximum \OIII luminosity observable from a PN), any time dependency is moot, as it can only lead to a reduction in luminosity. 

We start with the following premises:
\begin{enumerate}
\item We adopt the \citet[][hereafter MB]{Miller-Bertolami2016} function for the relation between the mass of a post-AGB star and its main-sequence progenitor (MB Figures 3 and 5).  Unlike some empirical IFMRs \citep[e.g.,][]{Cummings+18, ElBadry+18}, the MB function is notably flat (and slightly non-monotonic) in the initial mass range between 1.5 and $2.0 M_{\odot}$.  For {  the fits expressed as equations (1) and (2) below}, we neglect the metallicity dependence of the MB models {  which is quite small,} and fit {  the combined set of} points with a low-order polynomial,
\begin{equation}
\begin{split}
M_f = 0.3544 + 0.285936  M_i & - 0.125253 M_i^2 \\ &\hskip-20pt + 0.020942 M_i^3
\end{split}
\label{eq:mi-mf}
\end{equation}
This fit, for which the residuals are $\sim0.01 M_\odot$ (with the $Z=0.010$, $M_i = 3.0$ point excluded), is shown in Figure~\ref{fig:Mi-Mf_fit}. 

\begin{figure}[t]
\includegraphics[width=0.45\textwidth]{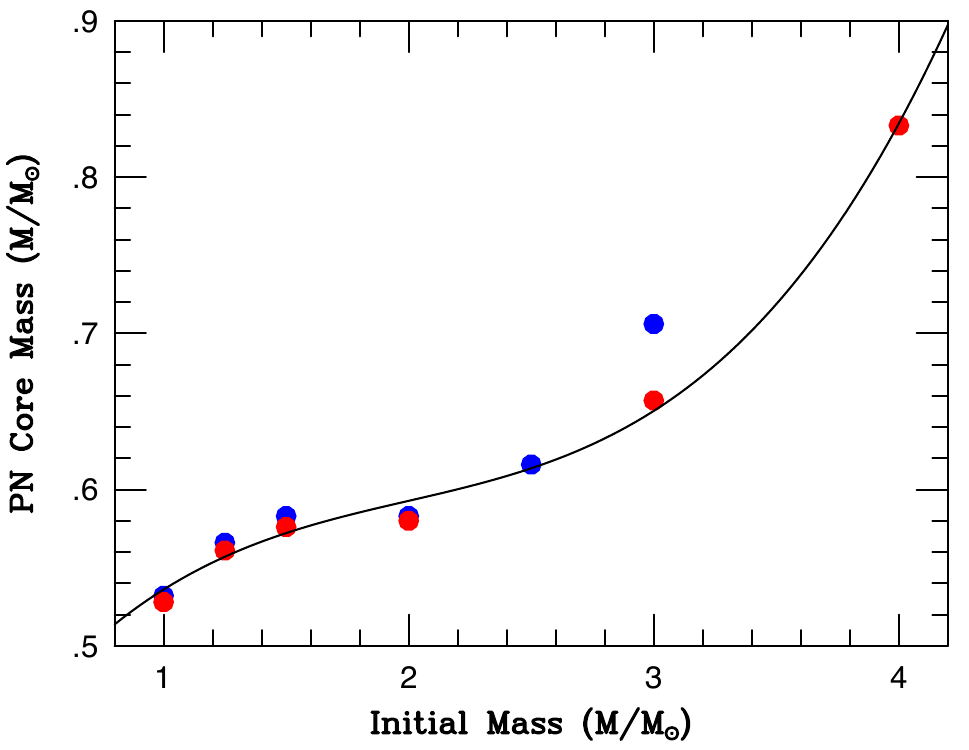}
\caption{Our polynomial fit to the MB initial-final mass relation. Data for $Z=0.010$ and $Z=0.020$ metallicities are shown as blue and red circles, respectively.  Our fit does not include the $Z=0.010$ point at $M_i=3.0$.}
\label{fig:Mi-Mf_fit}
\end{figure}

\item Similarly, we adopt the MB function relating the final mass of a star to its maximum bolometric luminosity during the post-AGB phase (MB Figure 11). As above, we approximate this $L_*$ value with a low-order monotonic polynomial,
\begin{equation}
\begin{split}
\log (L_*/L_\odot) = -70.6366+392.76992 M_f - \\ 772.82597 M_f^2 +672.848395 M_f^3 - \\ 217.746077 M_f^4
\end{split}
\label{eq:logL-mf}
\end{equation}

This fit is shown in Figure~\ref{fig:Lstar-Mf_fit}.

\begin{figure}[t]
\includegraphics[width=0.45\textwidth]{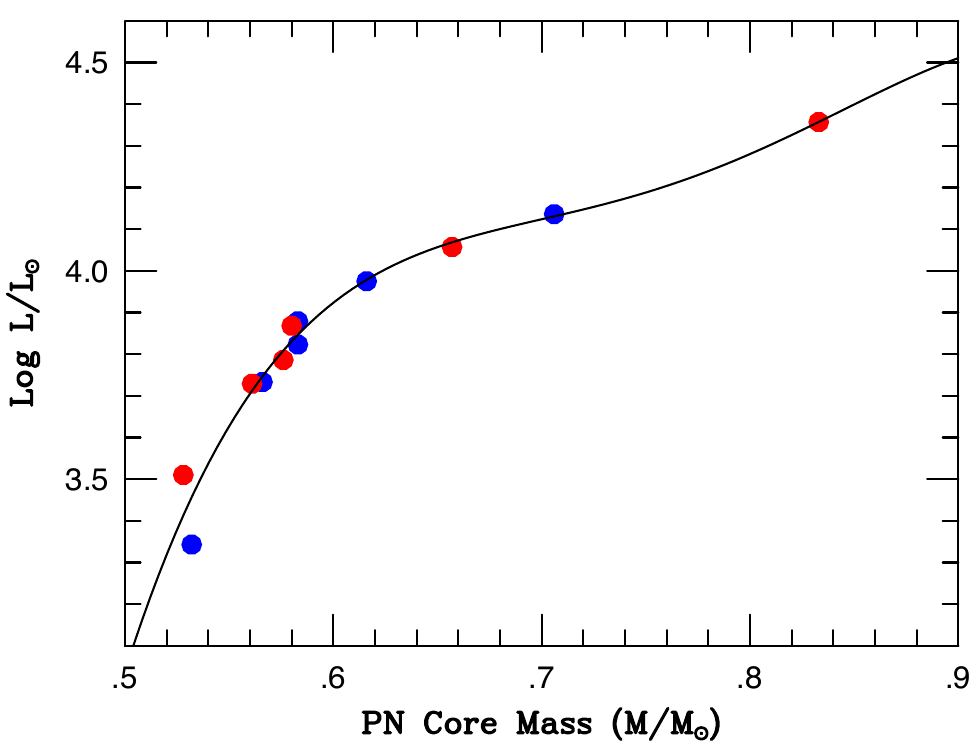}
\caption{Our polynomial fit for the variation of MB's central star luminosity versus core mass. Data for metallicities $Z=0.010$ and $Z=0.020$ are shown as blue and red circles, respectively. }
\label{fig:Lstar-Mf_fit}
\end{figure}

\item We assume that a PN can convert no more than 13\% of its central star's total luminosity into \OIII $\lambda 5007$ emission.  Photoionization models for the evolution of bright PNe generally predict a maximum conversion efficiency slightly greater than 12\%, with a small (roughly square root) dependence on metallicity \citep{Dopita+92, Schonberner+10, Gesicki+18}.  Figure~\ref{fig:efficiency} supports this number.  In the figure, the dereddened \OIII $\lambda 5007$ luminosities of the PNe analyzed in this paper are compared to their central stars' inferred bolometric luminosities.  While there is a noticeable decline in the efficiency of \OIII emission for objects fainter than $M_{5007} \sim -3.8$, at brighter magnitudes, the measurements are quite consistent, with a {  mean efficiency of $\sim 11.5\%$ and a scatter of $\sim 2.2\%$ about this mean.}  Since the four PNe with independent measurements suggest that the internal errors on the efficiencies are $\sim 2\%$, we can safely adopt 13\% as the maximum efficiency of \OIII $\lambda 5007$ production for \OIII-bright objects.

\item We use equations (1) and (2) to calculate the core masses and stellar luminosities of hypothetical PNe produced by populations with a given turnoff mass.  We then use step (3) to derive the maximum \OIII luminosity produced by these hypothetical PNe.

\item We use the data described in Section~\ref{sec:data} to derive a relationship between core mass and circumnebular extinction for the \OIII-bright PNe of the LMC and M31.  We then use this relationship to attenuate the \OIII luminosities of the hypothetical PNe to infer their observed brightnesses.
\end{enumerate}

\begin{figure}[t]
\includegraphics[width=0.45\textwidth]{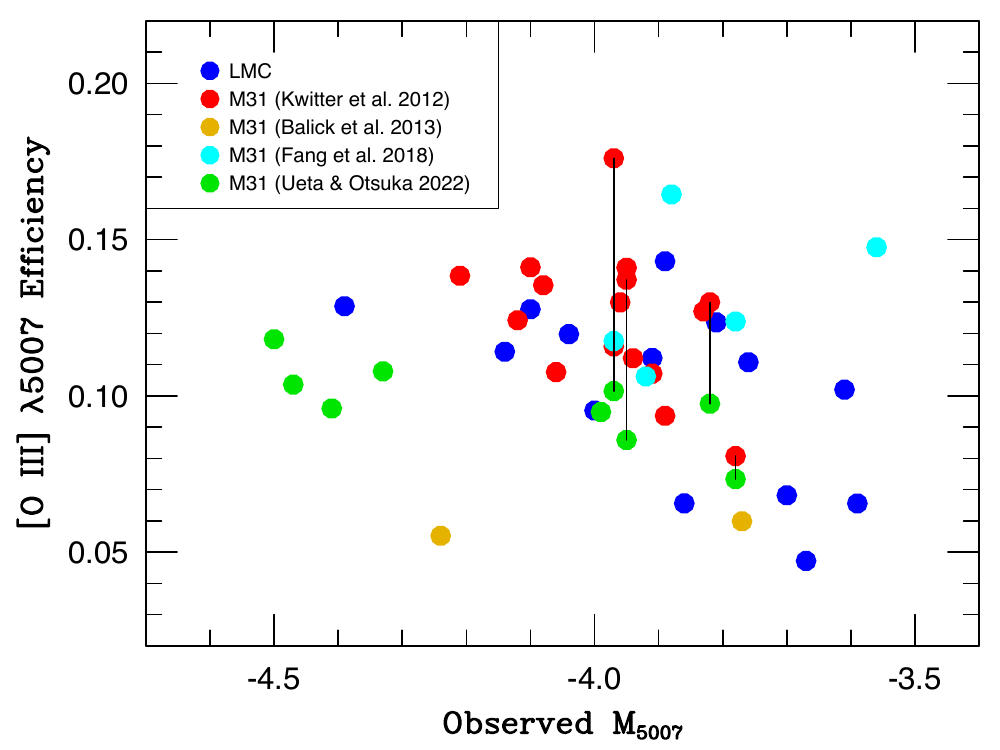}
\caption{The stellar luminosity to \OIII luminosity conversion efficiency for the PNe used in this study. The blue dots are the LMC PN measurements from \citet{Dopita+91a, Dopita+91b} and \citet{Dopita+97}, the red points represent the M31 PNe of \citet{Kwitter+12}, the two gold points show the metal-rich \citet{Balick+13} objects, the cyan dots are from \citet{Fang+18} and the green points are derived from the \citet{Ueta+22} reanalysis of the \citet{Galera+22} data. Four objects are common to the \citet{Kwitter+12} and \citet{Ueta+22} samples; these data are connected by vertical lines.  For PNe brighter than $M_{5007} \sim -3.8$, the mean and {  median efficiencies are essentially identical  ($\sim 11.5\%$), and the dispersion in the distribution (2.2\%)} is similar to that found from the PNe with repeat observations.}
\label{fig:efficiency}
\end{figure}

\begin{figure*}[t]
\includegraphics[width=0.95\textwidth]{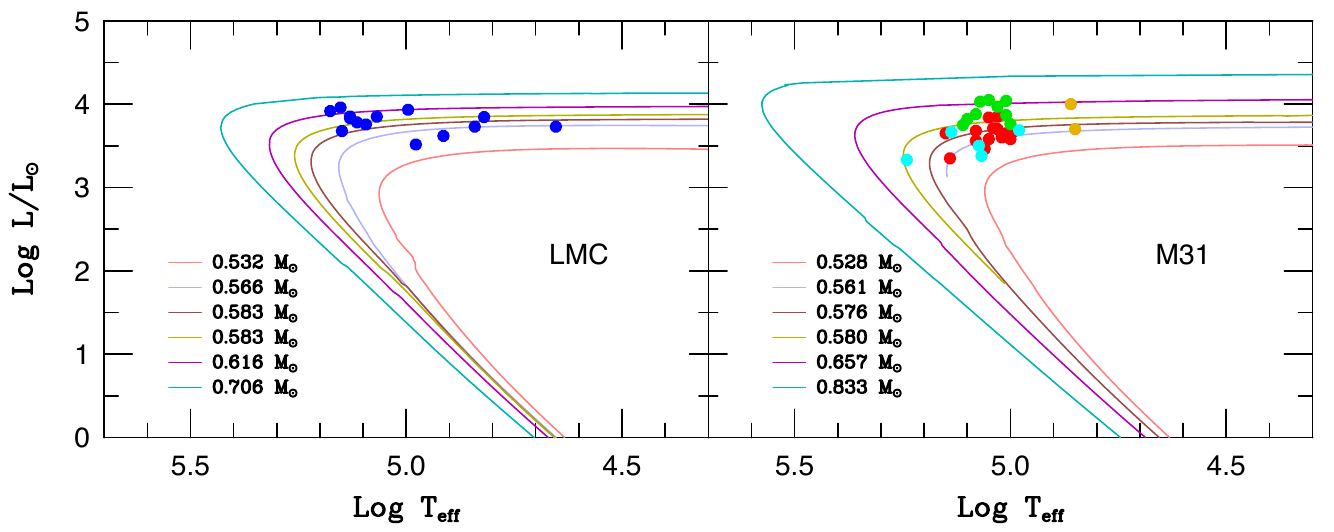}
\caption{The HR diagram positions of our LMC and M31 PN central stars compared to the MB post-AGB evolutionary tracks.  The left panel shows the LMC data and the MB models for $Z = 0.01$ stars; the right panel performs the same comparison for the M31 PNe and the MB $Z = 0.02$ tracks.  The color scheme is the same as that shown in Figure~\ref{fig:efficiency}. All the stars are still fusing hydrogen on their last pass across the HR diagram. {  There is an apparent, but not statistically significant, trend (Pearson $r \sim 0.34$; Spearman $\rho \sim 0.47$) in the LMC data where higher temperature stars have higher luminosities. This is likely a consequence of the small data set.}
}
\label{fig:hrdiagram}
\end{figure*}

The most critical aspect of this analysis is determining the relationship between the circumnebular extinction associated with an \OIII-bright PN and the object's core mass. To do this, we place each central star on the HR diagram using its ($\log T_{\rm eff}, \log L$) value, as derived from the authors' photoionization models.  We then infer the star's core mass by interpolating its HR-diagram position in MB's grid of post-AGB evolutionary tracks\null.  This 2-D interpolation is relatively straightforward since, as illustrated in Figure~\ref{fig:hrdiagram}, the vast majority of central stars are still undergoing their final, near-constant luminosity crossing of the HR diagram.\footnote{Our core-mass estimates supersede those published in articles that predate the MB calculations.} (In practice, our algorithm rotates the HR diagram until MB's post-AGB evolutionary tracks are single-valued in $x$.  The PN core masses are then found using simple 1-D interpolation in $y$.)  Finally, following \citet{Ciardullo+99} we regress the PNe's nebular extinctions against their calculated core masses. This relation, coupled with a \citet{Cardelli+89} extinction law with $R_V = 3.1$, allows us to calculate the expected amount of \OIII circumnebular extinction as a function of PN core mass.

\section{The Relationship Between Core Mass and Circumnebular Extinction}
\label{sec:extinction-coremass}

\begin{figure*}[t]
\includegraphics[width=\textwidth]{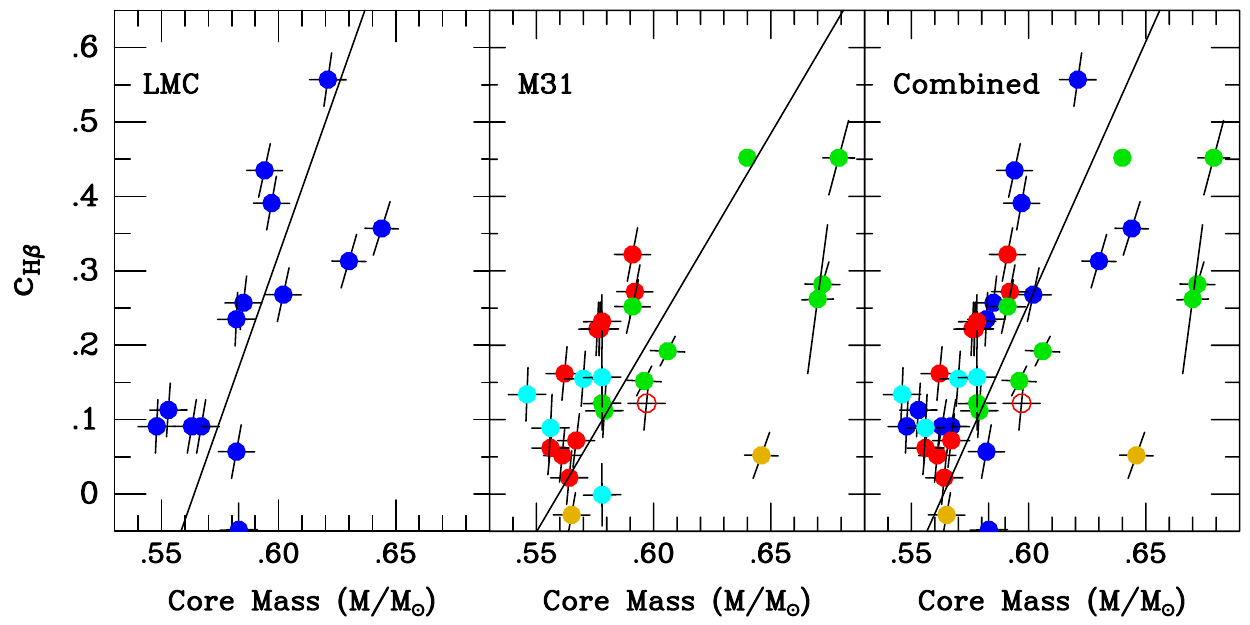}
\caption{
A comparison of circumnebular extinction versus derived PN core mass for LMC and M31 PNe within one mag of $M^*$.  The color scheme is that defined in Figure~\ref{fig:efficiency}.  The solid lines show our orthogonal regression fits that take into account the correlated errors between the two variables \citep{York+04}.  The open circle denotes a point that has been excluded from the fits because of the discrepancy between its plotted and tabulated position in \citet{Kwitter+12}.  There is a clear correlation between the two quantities. Even if our estimates for the systematic component of the errors were increased by a factor of $\sim 3$, the result would be qualitatively the same:  PNe with higher mass and, therefore, higher-luminosity central stars are more heavily reddened. {  As expected, there are a few slightly negative values of $c(H\beta)$; these are caused by measurement errors and the subtraction of Milky Way extinction.
}}
\label{fig:correlation}
\end{figure*}

Figure~\ref{fig:correlation} displays the extinction-core mass relation derived for the PNe of the LMC and M31. Before interpreting the figure, we must discuss the uncertainties associated with the measurements of the two variables.  {  To facilitate this analysis, we temporarily set aside any uncertainty associated with foreground extinction within the program galaxies.}

{ 
\subsection{Errors in Extinction}
}

The errors in $c_{{\rm H}\beta}$ come from comparing the PNe's observed Balmer decrements with expectations from low-density plasmas \citep[e.g.,][]{Pengelly64, Brocklehurst71, Storey+95}.  For the \citet{Ueta+22} PNe, these errors are quoted explicitly in their Table~1; for the \citet{Balick+13} and \citet{Fang+18} objects, $\sigma(c_{{\rm H}\beta})$ can be found by propagating the quoted uncertainties on each emission line strength (and assuming the error on the intrinsic value calculated from the modeling is negligible).  For the remaining objects, no information on the $c_{{\rm H}\beta}$ errors is given.  For these PNe, we assume an uncertainty of 0.036~dex, which is the median of the \citet{Balick+13}, \citet{Fang+18}, and \citet{Ueta+22} errors.   A variance test shows that if we treat the highly discrepant PN Merrett 2860 as an outlier, then the scatter in $c_{{\rm H}\beta}$ for the remaining three PNe with multiple measurements is consistent with this estimate.  

{ 
\subsection{Errors in Core Mass}
}

Next, we need to consider the random error on our core masses.  As described above, these come from interpolating the star's photoionization model-based $\log T_{\rm eff}$ and $\log L$ values in the grid of MB evolutionary tracks.  None of the papers give errors for the central star parameters, but based on our experience with photoionization models and the various trade-offs between the fitted variables, we believe that $\sigma(M_c) \sim 0.008$ is a reasonable estimate for the random component of the core mass error.\footnote{ {Since our targets for analysis are among the brightest PNe in the galaxy, they are most likely optically thick to ionizing radiation. In these types of objects, \citet{Kwitter+12} derive central star luminosities (and hence masses) primarily via the extinction-corrected fluxes of the brightest Balmer lines, which have a quoted error of $10-15\%$.  When we increment the reported stellar luminosities by this amount, we find that the mean mass of a central star increases by $\sim 0.008 M_\odot$.}}  Note that this error does not include the systematic component associated with our choice of post-AGB evolutionary tracks, {  nor does it reflect variations in derived core masses based on significantly discrepant emission-line observations.}

{ 
\subsection{The Impact of Correlated Errors on the Extinction-Core Mass Relation}
}

Finally, we must consider the likely presence of a correlated error between extinction and core mass. This systematic comes about because any overestimate of extinction may cause the photoionization models to overestimate the total luminosity of line emission from the nebula and, therefore, (through energy balance) the total luminosity of the exciting star. This error can then propagate directly into the core mass values.

To estimate the amplitude of this correlated error, we combine our estimate of $\sigma(c_{{\rm H}\beta})$ with an $R_V = 3.1$ \citet{Cardelli+89} extinction law and calculate the effect that a given error in H$\beta$ extinction has on the total amount of \OIII luminosity being emitted.  Because all the objects in our sample are dominated by \OIII emission, this effect is always very close to $0.96 \, c_{{\rm H}\beta}$.

At the same time, we estimate the (very small) effect an error in $c_{{\rm H}\beta}$ has on $\log T_{\rm eff}$ using the \ion{He}{1} $\lambda 4686$/H$\beta$ ratio and the cross-over method \citep{Kaler+89}.  Although this technique only works for optically thick nebulae with hot ($T > 80,300$~K) central stars, we expect that most of the PNe in the top $\sim 1$~mag of the \OIII PNLF should at least come close to satisfying these conditions. To wit: if Lyman continuum photons are leaking out into interstellar space, then the efficiency of \OIII production would necessarily decrease, leading to fainter \OIII magnitudes.  Similarly, PNe with cooler central stars will have a larger fraction of oxygen in the singly ionized state, again suppressing the efficiency of \OIII emission.  But most importantly, even if a PN is not optically thick and/or has a relatively cool central star, the core's location on the HR diagram should guarantee that it is in its final, near-constant-luminosity trek across the HR diagram (see Fig.~\ref{fig:hrdiagram}); for these objects, any error in effective temperature has little, if any, effect on the measurement of core mass.  

Finally, we use the error vectors calculated above to find new values for $\log L$ and $\log T_{\rm eff}$ and repeat Step 6 to find the expected change in core mass due to our hypothesized error in extinction.  The random and systematic components are then combined to create a covariance matrix for each PNe.  These matrices are represented by the error vectors shown in Figure~\ref{fig:correlation}.

From the figure, it is clear that our derived $c_{{\rm H}\beta}$-core mass relation contains a significant amount of scatter; this comes both from measurement error and lack of knowledge about foreground extinction affecting each PN\null.  Nevertheless, the datasets for both galaxies follow the overall trend found by \citet{Ciardullo+99} and \citet{Kwitter+12}:  there is a correlation between the circumnebular extinction of \OIII-bright PNe and their core mass.  Interestingly, although the LMC PNe have half the metallicity of their M31 counterparts (which comes principally from the outer disk), the extinction-core mass relation for the two systems are similar.  If we apply the orthogonal-regression (OR) algorithm of \citet{York+04}, which fits heteroskedastic data in the presence of correlated errors, we obtain slopes of $8.9 \pm 2.6$ for the LMC, $5.3 \pm 1.7$ for M31, and $7.0 \pm 1.8$ for the combined dataset.  

We note that simpler algorithms, such as ordinary least squares (OLS), yield shallower slopes for the $c_{{\rm H}\beta}$-core mass relation, with values generally between $\sim 3$ and $\sim 5$.  This is to be expected: while OLS assumes the observed slope is entirely due to the relationship between the two variables, the \citet{York+04} OR algorithm treats the slope as being due to two components, one intrinsic to the data and the other a product of the correlated errors.  But regardless of the type of least squares algorithm used, the key result of the analysis is the same:  the correlation between $c_{{\rm H}\beta}$-core mass is real.  Even if our best estimates for the systematic component of the errors were increased by a factor of $\sim 3$, there would be no qualitative difference in the result.\footnote{Our LMC figure has far less scatter than the equivalent version in \citet{Ciardullo+99}.  Unlike the previous work, the present sample is restricted to PNe within the top one mag of the \OIII luminosity function.  This more appropriate selection criterion results in a more uniform PN population.}  

Table~\ref{tab:slopes} summarizes our best-fit relations.  Given the heterogeneous nature of the data and our lack of knowledge about true errors on the core masses, the formal uncertainties for the fitted slopes and intercepts may be underestimates.  In fact, depending on the assumptions used for our fitting procedure, our analysis admits a wide range of slopes, from $b \sim 4$ to $b \sim 9$.  Similarly, because the intercept values are also uncertain, we include an extra column in the table labeled ``Intercept*'', which given the best-fit slope, is the intercept needed to ensure that the peak value of $M_{5007}$ matches the observed value of $M^*$.  In all cases, the offset between Intercept* and the formal best-fit intercept is quite small, i.e., just a few percent of the uncertainty.

Of course, since the composition and distribution of the AGB dust ejecta is expected to vary with stellar mass, due to the changing physics of dredge-up and time since the AGB \citep[e.g.,][]{Karakas+2016, Dellagli+23}, we might expect the actual relation to have features (e.g., non-linearities and discontinuities) that are hidden within the uncertainties of the measurements and the limited PN samples. Nevertheless, for purposes of this study, a simple linear relation will do, and the large positive slope suggests an answer to the long-standing puzzle of the constancy of the PNLF cutoff.

\begin{deluxetable}{lccc}[t]
\tablecaption{Orthogonal Regression of $c_{{\rm H}\beta}$ with Core Mass \label{tab:slopes} }
\tablehead{
\colhead{PN Sample} &\colhead{Slope} &\colhead{Intercept} &\colhead{Intercept$^*$}}
\decimals
\startdata
LMC OLS  &$4.18\pm0.93$ & $-2.24\pm0.54$ & $-2.27$ \\
M31 OLS  &$2.09\pm0.55$ & $-1.06\pm0.32$ & $-0.98$ \\
Combined OLS  &$2.53\pm0.53$\ & $-1.30\pm0.30$ & $-1.26$ \\
LMC OR &$8.92 \pm 2.62$ &$-5.03 \pm 1.55$ & $-4.99$ \\
M31 OR &$5.35 \pm 1.65$ &$-3.00 \pm 0.97$ & $-2.97$ \\
Combined OR &$7.05 \pm 1.77$ &$-3.97\pm1.04$ & $-3.94$ \\
Optimized (see text) & 5.4 & $-3.0$ & \dots \\
\enddata
\tablenotetext{*}{The intercept value required to yield a maximum $M_{5007}$ equal to $M^*$. The difference from this value and the best-fit intercept is well within the quoted uncertainty. }
\end{deluxetable}

\subsection{An Independent Check of Circumnebular Extinction:  Type I PNe}
\label{sec:Type1}

Because our explanation for the invariance of the PNLF cutoff hinges on the systematic behavior of circumnebular extinction, it would be useful to have an independent way of verifying that the results of Figure~\ref{fig:correlation} are not simply due to correlated errors.  The systematics of PN chemical abundances provides such a test.   

Without circumnebular extinction, the brightest PNe within a star-forming galaxy should be those that evolve from the highest mass ($3 \lesssim M/M_{\odot} \lesssim 8$) PN progenitors. Such stars have a unique signature: unlike their lower-mass counterparts, high-mass stars undergo a third dredge-up and hot-bottom burning while on the AGB \citep[e.g.,][]{Herwig2005}.  As a result, PNe which evolve from these objects will be enhanced in nitrogen and helium.  In other words, {  the most luminous PNe observed in a galaxy should be Type~I planetaries.}

As defined by \citet{Peimbert+TP83}\footnote{See also \citet{Torres-Peimbert+1997} for a discussion of alternative definitions of Type~I PNe. Under some of these definitions, several of the M31 PNe marginally satisfy one of the Type~I criteria, i.e., are slightly overabundant in either helium or nitrogen.}, Type~I PNe are nebulae with $\log {\rm N/O} \ge -0.3$ and He/H $> 0.125$.  None of the 28 M31 PNe analyzed by \citet{Kwitter+12}, \citet{Balick+13}, \citet{Fang+18}, and \citet{Ueta+22} satisfy these criteria.  Similarly, only one of the 14 LMC PNe included in our analysis has a Type~I classification (SMP 47). Thus, at best, only $\sim 3\%$ of PNe in the top magnitude of the \OIII luminosity function are formed from high-mass stars.

We can compare this number to the results of \citet{Jacoby+99}, who analyzed the abundances and central star properties of PNe in the bulge and disk of M31.  These authors found that over a quarter of their sample (4 out of 15) satisfy the Type~I criteria, but most interestingly, all the Type~I objects are two to three magnitudes fainter than $M^*$.  This result supports the notion that the PNe which populate the bright end of the luminosity function do not evolve from high-mass progenitors.

{  
We note that circumnebular extinction may not solely be responsible for the lack of Type~I planetaries in the top magnitude of the PNLF; because of their shorter post-AGB lifetimes, the chance of observing a Type~I PN is smaller than that of finding PNe from lower-mass stars.  Unfortunately, at present, the relative importance of the two mechanisms is difficult to quantify. 

We can calculate the relative rates of PN formation against their progenitor masses from the application of the fuel-consumption theorem \citep{Renzini+86}. But any calculation for the number of PNe expected to be seen in a survey of the top $\sim 1$~mag of the PNLF would not only require detailed modeling of the time evolution of the PNe's central stars, their nebulae, but also the formation/destruction of their dust and its spatial distribution.  Such calculations are beyond the scope of this paper.  

Instead, we compare the fraction of Type~I PNe in the Galaxy to the fraction seen in our sample of LMC/M31 objects.  There are $\sim 3500$~confirmed (and unconfirmed) PNe cataloged in the HASH database of Milky Way planetary nebulae \citep{Parker+16}, and roughly 2\% are classified as Type~I \citep{Phillips2004}. This is similar fraction found for the objects listed in Table~\ref{tab:pnlist} (one out of 42).  

So, although we cannot quantify the relative importance of either of the two mechanisms, we see no clear argument that the short visibility time for Type~I PNe is dominating. Furthermore, Figure~\ref{fig:M5007_cutoff} demonstrates that rapid evolution does not dramatically suppress the presence of stars with progenitor masses up to $\sim3.3M_\odot$ in the top mag of the luminosity function.
}

\section{The Effect of Dust on the PNLF}
\label{sec:results}

\begin{figure*}
\includegraphics[width=0.9\textwidth]{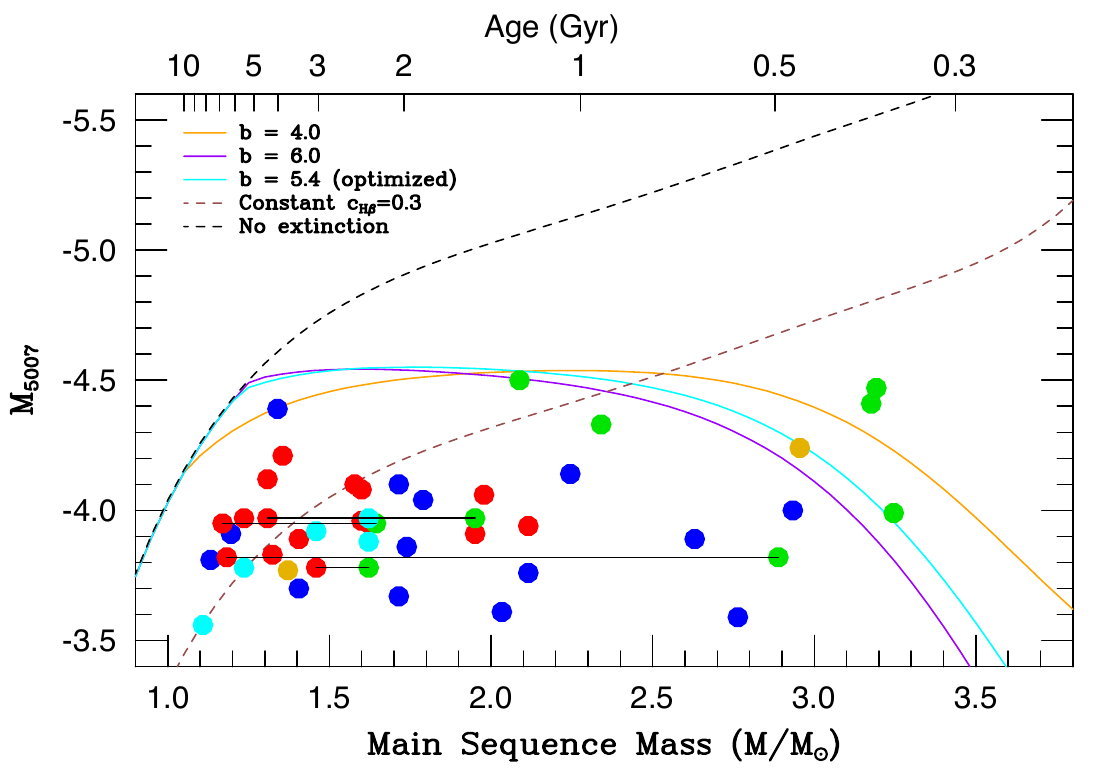}
\caption{Curves showing the behavior of $M^*$ versus progenitor star mass (and age) for three different slopes of the $c_{{\rm H}\beta}$ vs.\ $M_{f}$ relation; the $b = 5.4$ slope produces the least variation of $M^*$  over population age.  Also shown are the curves for where the extinction is assumed to be independent of stellar mass ($c_{{\rm H}\beta} = 0.3$) and a zero extinction model.  The points show the inferred locations of LMC and M31 PNe, with the color scheme defined in Figure~\ref{fig:efficiency}. The four objects common to the \citet{Kwitter+12} and \citet{Ueta+22} samples are connected with horizontal lines.  The data do not explain why \OIII-bright PNe are observed in old populations, but they do demonstrate how all populations younger than $\sim 5$~Gyr can have virtually the same value of $M^*$. Without the action of circumnebular dust, the PNe from high-mass progenitors would have absolute \OIII\ magnitudes much brighter than those observed across nearly 100 galaxies observed to date.
\label{fig:M5007_cutoff}}
 \end{figure*}

Once we adopt a slope for the extinction core-mass relation, it is relatively straightforward to use the steps described in \S\ref{sec:approach} to estimate a PN's largest observable \OIII luminosity as a function of its progenitor mass (or age).  The results of this analysis are shown in Figure~\ref{fig:M5007_cutoff}. The figure has a number of interesting properties.  

The first is the behavior of the curve when circumnebular extinction is ignored or simply assumed to be independent of progenitor mass.  Not surprisingly, under these conditions, the absolute magnitude of the PNLF cutoff becomes a sensitive function of progenitor mass or, equivalently, population age. This is the behavior reported by the modeling efforts of \citet{Marigo+04}.

A second feature of the figure is the fading of $M^*$ in the oldest stellar populations.  This is the issue considered by \citet{Ciardullo+05}: under most IFMRs, when a stellar population becomes old enough, the central stars formed from single-star stellar evolution do not produce enough luminosity to create $M^*$ planetaries.  While the faster-evolving post-AGB models of \citet{Miller-Bertolami2016} help mitigate the problem, the issue is still unsolved.  Possible solutions include alternative forms of stellar evolution, such as that which creates blue straggler stars \citep{Ciardullo+05}, a flatter IFMR at low stellar masses \citep[e.g.,][]{Cummings+18}, or a non-negligible amount of dispersion in the IFMR (see Section~\ref{sec:discussion}).

The third and most important property of Figure~\ref{fig:M5007_cutoff} is the behavior of the models where the slope of the extinction-core mass relation is between 4 and 6.  Under these conditions, the cutoff magnitude varies only slightly over a wide range of progenitor mass.  

{  We also consider an ``optimized'' case that maximizes the age range over which the function is more-or-less constant.} Here, the slope is 5.4 and the intercept is $-3.0$, and the cutoff magnitude changes by less than $\pm 0.05$ mag between $1.4 < M/M_{\odot} < 2.7$ and by less than $\pm 0.10$ mag between $1.25 < M/M_{\odot} < 3.1$.  The latter set of limits corresponds to stellar population ages between $\sim 0.4$ and $\sim 5$~Gyr.

The behavior of the curves plotted in Figure~\ref{fig:M5007_cutoff} demonstrates that circumnebular extinction {  is a likely candidate for} the physical mechanism behind the near universality of the PNLF cutoff.  For slopes of the extinction core-mass relation between $\sim 4$ and $\sim 6$, a wide range of population ages produce PNe with approximately the same value of $M^*$. For the very youngest populations, the curves of Figure~\ref{fig:M5007_cutoff} turn downward, reflecting the fact that PNe from high-mass progenitors do not populate the top $\sim 1$~mag of the PNLF\null.  This is consistent with the results of Section~\ref{sec:extinction-coremass} and the data points displayed in Figure~\ref{fig:M5007_cutoff}.

The issue of the turnover for very old stellar populations is more complicated.  As mentioned above, with our present understanding of stellar evolution and post-AGB star physics, the decline in $M^*$ for very old populations is unavoidable.  Low-mass main sequence stars create low-mass PN central stars, and these stars simply are not bright enough to produce $\sim 620 L_{\odot}$ of \OIII $\lambda 5007$ to be emitted from a nebula.  However, if old populations could build some high-mass cores --- either through binary evolution/mergers or stochastic processes, then Figure~\ref{fig:M5007_cutoff} would explain why the PNLF is such a robust standard candle.  

Details of the shape of the PNLF are known to vary from galaxy to galaxy \citep[e.g.,][]{Jacoby+02, Reid+10, Ciardullo2010, Bhattacharya+21}.  Similarly, the bolometric-luminosity specific number of \OIII-bright PNe is also population dependent \citep{Ciardullo+05, Ciardullo2010}.  But the PNLF distance to a galaxy is defined by the very brightest objects:  as long as a galaxy can build the cores typically associated with the end products of 1.3 to $2.7 M_{\odot}$ main sequence stars,  the actions of circumnebular dust will ensure that the observed cutoff in the luminosity function remains very near $M_{5007} \sim -4.5$. 
 
\section{Discussion}
\label{sec:discussion}

\subsection{Successes of the Demonstration}
\label{subsec:success}

\citet{Gesicki+18} showed that one can derive peak \OIII luminosities for PNe that meet or exceed the observed value of $M_{5007} \sim -4.5$ by simulating PNLFs for stellar populations younger than $\sim 3$~Gyr (i.e., progenitor masses greater than $\sim 1.5 M_{\odot}$).  Their analysis tracks stellar evolution from the main sequence through the post-AGB phase, follows the photoionization history of the post-AGB stars' nebulae, and includes a prescription for extinction that involves the progenitor AGB star's superwind mass-loss rate and velocity.

This paper does not attempt such comprehensive modeling; instead, it focuses narrowly on assessing the maximum \OIII luminosity achievable for post-AGB stars of various masses, using knowledge of the maximum efficiency for converting stellar luminosity into \OIII $\lambda 5007$ emission and an empirical prescription for the behavior of circumnebular extinction with core mass.  With this approach, the uncertainties associated with nebular physics and galactic star formation histories are irrelevant.  While our analysis prevents us from deriving an actual luminosity function, it does generate guidelines for future analyses.

Ultimately, our results are not so different from those of \citet{Gesicki+18}. However, our analysis shows that not only is dust necessary for modeling the PNLF, but it is perhaps \textit{the} driving force behind the invariance of the PNLF's bright-end cutoff.  Moreover, this conclusion follows directly from observations rather than any attempted mating of stellar evolution and nebular modeling.  As long as a stellar system has stars with main sequence masses of  $1.25 M_{\odot}$ or greater (younger than $\sim5$ Gyrs), dust ensures that the PNLF's high-luminosity cutoff will yield a distance that is accurate to $<2.5\%$, provided the observed PN sample is sufficiently large to populate the brightest $\sim 1$~mag of the PNLF with $ \gtrsim 30$~objects.

\subsection{The Turn Over in Young Stellar Populations}
\label{subsection:young_age_failure}

As discussed in  Section~\ref{sec:extinction-coremass}, PNe which evolve from stars with initial masses $M \gtrsim 3 M_{\odot}$ generally do not populate the bright-end of the PNLF\null.  This is illustrated by the solid curves of Figure~\ref{fig:M5007_cutoff}, all of which display a turn-over at high masses. While interesting in theory, this feature has no practical effect on PNLF distances.  Stars with initial masses of $M \gtrsim 2 M_{\odot}$ have relatively short lifetimes (less than $\sim 1$~Gyr), and there are no large galaxies in the local universe that are composed solely of stars this young.  Even in the most vigorous star-forming galaxies, the PNLF cutoff will be defined by the objects that have evolved from the underlying older populations.  The existence of this fading for young populations is, therefore, a moot issue for distance determination.

\subsection{Failure in Old Stellar Populations}
\label{subsection:high_age_failure}

A more serious concern is the unobserved fading of the PNLF in old stellar populations. Elliptical galaxies are dominated by stars with ages $\gtrsim 6$~Gyr \citep[e.g.,][]{Trager+2000b, Zibetti+20, Parikh+24}; thus, according to Figure~\ref{fig:M5007_cutoff}, their PNLF cutoffs should be $\gtrsim 0.5$ mag fainter than those seen in spiral galaxies.\footnote{{  While a moderate fraction of field ellipticals show signs of recent star formation, the phenomenon is rare in groups and clusters, where most PNLF measurements have been made \citep{Paspaliaris+23}.  In any case, since the amount of luminosity in this ``young'' component is usually small, the luminosity-specific stellar evolutionary flux \citep{Renzini+86, Buzzoni+06} predicts that it would not be sufficient to populate the bright-end of the PNLF.\null}}  Deviations of this magnitude would produce PNLF distances that are $\sim 25\%$ greater than those found by other methods.  This is not the case: if there is any systematic bias to the PNLF, it is towards too small a distance for old populations, and its amplitude is less than $\sim 10\%$ \citep{Ferrarese+00, Feldmeier+07}.

This mismatch between the predicted and observed PN luminosities for Pop~II systems has been discussed several times in the literature \citep[e.g.,][]{Jacoby+89, Marigo+04, Ciardullo+05, Gesicki+18, Yao+23} along with various possible solutions: mass enhancement through binary interactions and stellar mergers \citep{Ciardullo+05}, the presence of a small population of intermediate-age stars whose signature is masked by red giants \citep{Jacoby1997, Ciardullo+05}, an incorrect reference PNLF \citep{Hartke+17}, contamination by Pop~I emission-line sources such as Wolf-Rayet nebulae \citep{Jacoby+2024}, the presence of interloping $z \sim 3.1$ Ly$\alpha$-emitting galaxies in the PN sample \citep{Kudritzki+00, Jacoby+2024}, or even contamination by post-AGB objects whose nebulae are powered by accretion onto companion white dwarfs \citep{Soker2006, Souropanis+23}.   To these, we add three additional possibilities:

\subsubsection{An overly steep initial-to-final mass relation.}  The MB models' prediction for the initial-to-final mass relation may be $\sim 0.03$ $M_{\odot}$ too low in the regime of interest ($M \lesssim 1.25 M_{\odot}$).  Although their IFMR is in good agreement with the Gaia-based analysis of substructure in the white-dwarf cooling sequence \citep{ElBadry+18}, the physics producing the relation is extremely complex, and there are few direct measurements for the masses of white dwarfs in clusters older than $\sim 5$~Gyr \citep{Kalirai+08}.

In fact, there are suggestions for a flatter IFMR in the literature.  The MB models predict white dwarf masses of $\sim 0.53 M_{\odot}$ for solar mass progenitors, yet the semi-empirical IFMR presented by \citet{Cummings+18} in their equations (1) and (4) yield core masses $\sim 0.03 M_{\odot}$ larger than this, i.e., $\sim 0.566 M_{\odot}$.  Similarly, in their investigation of the physics of the IFMR, \citet{Addari+24} found a relationship wherein $1 M_{\odot}$ stars produced $0.551 M_{\odot}$ cores.  When combined with Fig~\ref{fig:Lstar-Mf_fit} above, either of these results would generate PNe with $M_{5007} \approx -4.5$ in 10 Gyr populations. 

\subsubsection{A scatter in the IFMR\null.}  The existence of an IFMR does not, in itself, require a unique relationship between the main-sequence mass of a star and its mass on the post-AGB\null. The empirical IFMRs derived from white dwarf measurements in star clusters typically exhibit a substantial amount of scatter \citep[e.g.,][]{Kalirai+08, Cummings+18, Canton+21}. Much of this scatter may be due to measurement error, but some intrinsic dispersion about a mean IFMR is not ruled out.  
    
Indeed, a number of mechanisms (e.g., stellar rotation, abundance variations, the presence of companion stars or planets, or even simple stochasticity driven by the physics of thermal pulses) may introduce scatter into the relation; if this occurs, even the oldest stellar systems may produce some cores massive enough to excite an $M^*$ planetary.  This possibility is supported by a number of observations, ranging from the $\sim 0.02 M_{\odot}$ dispersion generally attributed to mass loss on the red giant branch \citep[e.g.,][]{Lee+94, Dotter2008, Catelan2009} to the scatter of white dwarf masses in star clusters \citep[e.g.,][]{Kalirai+08, Cummings+18}.  
       
If there is a stochastic element to the IMFR, then the PNLF cutoff could maintain its luminosity even in the oldest of stellar populations. However, this stochasticity would also have a secondary effect: it would cause the production rate of \OIII-bright PNe to decline with age, as fewer and fewer stars would achieve the necessary core mass to produce $M^*$ PNe.  

To illustrate this, consider an MB-like IFMR but with an intrinsic Gaussian dispersion of $\sigma = 0.022$ in the final white dwarf mass.  According to equation~(2) and our assumed value of 13\% for the maximum efficiency of \OIII production, the creation of an $M^*$ PN requires a central star of $M \gtrsim 0.564 M_{\odot}$.  Under our Gaussian dispersion model, 50\% of stars with an initial mass of $1.35 M_{\odot}$ would create a PN capable of emitting $\sim 620 L_{\odot}$ in the \OIII $\lambda 5007$ line, but only 10\% of cores from $1 M_{\odot}$ progenitors would meet this threshold.  This factor of $\sim 5$ decline in the luminosity-specific number densities of \OIII-bright PNe is exactly the difference between that seen in ``blue-ish'' E/S0 galaxies, such as NGC\,5102, NGC\,5128, and NGC\,1316, and red giant ellipticals, such NGC\,1399 and NGC\,1404 \citep{Coccato+09, Ciardullo2010, Jacoby+2024}. {  This factor cannot be attributed to the ratios of young versus old populations because the luminosity-specific stellar death rates (and hence, the PN production rates) are nearly constant across populations \citep{Renzini+86}.}
    
\subsubsection{A high \OIII conversion efficiency}  The efficiency for the conversion of bolometric stellar luminosity to \OIII $\lambda 5007$ emission may be significantly higher than our assumed value of $13\%$. This possibility seems unlikely:  no modeling effort has exceeded this threshold \citep[e.g.,][]{Dopita+92, Schonberner+10}, and Figure~\ref{fig:efficiency} presents empirical evidence for a $\sim13\%$ cap. However, larger samples of objects are needed to determine the full distribution of efficiencies. 

\subsection{The Scatter in the Extinction-Core Mass Relation}
\label{sec:scatter}

As illustrated in Figure~\ref{fig:correlation}, the extinction-core mass relation exhibits a large amount of scatter.  {  Some of this scatter is likely due to internal extinction in the host galaxies; the slightly larger scatter in the M31 dataset lends some support for this interpretation, although the more heterogeneous nature of the data set may also be responsible.}  

Another factor is undoubtedly measurement error, as the process of using a limited number of optical (and in some cases, near-IR) emission lines to fix a central star's position on the HR diagram is quite complex.  There are a large number of trade-offs and degeneracies associated with the process, and the result can be non-unique solutions for the stellar luminosity and temperature.   

It is also likely that much of the dispersion is intrinsic.  The distribution of dust in bipolar PNe is notoriously asymmetric, and even in roughly spherical nebulae, walls, ridges, and filaments of extinction are easily seen. {  (For excellent examples, see the in-depth study of NGC\,7027 by \citet{MoragaBaez+23} and the set of PN extinction maps created by \citet{Pignata+24}.)} Thus, the value of $c_{{\rm H}\beta}$ for distant unresolved PNe is likely sight-line dependent.  This effect alone would introduce scatter into Figure~\ref{fig:correlation}.

Finally, the effects of extinction on the \OIII luminosity of planetary nebulae are undoubtedly time-dependent, as dust may get destroyed or dissipate.  In this study, we attempted to create as homogeneous a PN sample as possible by restricting our analysis to objects within 1~mag of the PNLF cutoff.  Yet, as Fig.~\ref{fig:efficiency} shows, there is a suggestion of evolution in our data, as the objects in the top 0.6~mag of the PNLF are more efficient at producing \OIII $\lambda 5007$ than their fainter counterparts.  This would also introduce intrinsic scatter into Figure~\ref{fig:correlation}.

If there is scatter in the extinction-core mass relation, then the shape of the brightest $\sim 0.5$~mag of the PNLF may largely be defined by stochastic effects associated with the dust distribution, such as viewing angle.  A much larger sample of PNe will be needed to test this hypothesis. 

\section{Conclusions}
\label{sec:conclusions}

When considering why the PNLF works as a distance indicator across the different stellar populations of a diverse set of galaxies \citep{Jacoby+2024}, two primary invariances must be understood: age and metallicity. The sensitivity to metallicity has been discussed in the literature and is reasonably well understood \citep[e.g.,][]{Dopita+92, Ciardullo+92}.  In brief, the combination of stellar and nebular physics mitigates the effect of metallicity on the PNLF so that changes in $M^*$ are only weakly dependent on a population's oxygen abundance.  In systems more metal-poor than [O/H] $\sim -0.3$, $M^*$ does fade \citep{Ciardullo+02}; however, the presence of a mass-metallicity relation for galaxies \citep[e.g.,][]{Tremonti+04} ensures that systems with such low abundances will have few if any PNe populating the brightest magnitude of the PNLF {  and thus, those galaxies are not applicable candidates for distance determinations anyway.} 

The PNLF cutoff may also fade in galaxies with super-solar abundances \citep{Dopita+92}, but since such systems always contain a mix of stellar populations, the effective value of $M^*$ will invariably be that defined by the brighter, more intermediate-metallicity stars.  Thus, abundance variations should play a very small role in determining the absolute magnitude of the PNLF cutoff in galaxies with enough PNe for a precision distance measurement. If a metallicity correction is needed, it can be applied via the use of a simple look-up table or formula  
\citep[e.g.,][]{Ciardullo+02}.  

The question of why the $M^*$ does not depend on population age is a more vexing issue and is one that has persisted for the 35-plus years that the PNLF has been in use.  The problem has two separate components.  The first is whether the post-AGB cores of old stellar populations can produce enough luminosity to excite $\sim 620 L_{\odot}$ of \OIII $\lambda 5007$ emission in planetary nebulae.  Although this paper does not address this problem directly, we do point out that a number of possible solutions do exist, including the introduction of a small dispersion into the mean IMFR relation.  Progress on this front requires a program that is aimed at not only improving white dwarf mass measurements in old clusters but also quantifying the intrinsic scatter in the results.   

The second component of the problem is explaining why the observed \OIII $\lambda 5007$ luminosities of PNe in large star-forming galaxies are limited to $L \lesssim 620 L_{\odot}$.  In this work, we argue that the driving mechanism behind this limit is circumnebular extinction.  Based on stellar evolution theory, there are several reasons to believe that PNe from higher-mass stars should have more circumnebular reddening than their lower-mass counterparts.  We show that, indeed, the available data confirms this hypothesis.  More importantly, we show that this relation can produce a behavior where the PNLF cutoff appears to be very nearly invariant over a wide range of stellar populations.  

The systematic behavior of circumstellar dust turns the PNLF's greatest limitation into an asset.  A high-quality PNLF distance requires that a galaxy contain $\gtrsim 30$~PNe in the top magnitude of the PNLF\null.  Effectively, this means the technique is restricted to moderately luminous galaxies, i.e., $M_V \lesssim -19.5$.  Since systems such as these always contain an underlying population of older, intermediate-metallicity stars, there will invariably be a reservoir of objects available to populate the brightest magnitude of the PNLF\null.  Therefore, the fading of $M^*$ for the highest mass PN cores is never seen.

Our analysis of the systematics of circumnebular extinction surrounding planetary nebulae is limited by the precision of the central star mass estimates, all of which come from photoionization modeling.  In particular, these data, which were derived by interpolating the inferred positions of the central stars in MB's grid of post-AGB evolutionary tracks, come from seven different papers and four different investigative teams.  Moreover, none of these works report uncertainties for the central stars' effective temperatures and luminosities. Thus, it is impossible to determine how much of the scatter seen in Figure~\ref{fig:correlation} is intrinsic.  Better constraints on the central star properties, perhaps via the inclusion of ultraviolet and infrared data in the photoionization models, would improve the situation and reduce the effect of correlated errors on the extinction-core mass relation.

\section{Acknowledgments} 

The authors wish to recognize the significant contributions of Karen Kwitter, who recently passed away, to the field of planetary nebulae. Her 2012 paper \citep{Kwitter+12} formed the critical foundation for this work, and her recent review article with Richard Henry \citep{Kwitter+Henry2022} will remain an invaluable reference for future research in the field.

{ 
We thank Martin Roth, Magda Arnaboldi, and Lucas Valenzuela for providing important suggestions and corrections to an earlier version of the paper. We especially thank the referee, Albert Zijlstra, for carefully reading the manuscript and offering critical recommendations for improvements.
}

The Institute for Gravitation and the Cosmos is supported by the Eberly College of Science and the Office of the Senior Vice President for Research at the Pennsylvania State University.

\clearpage 

\bibliography{MUSE-PNLF_II}{}
\bibliographystyle{aasjournal}

\end{document}